\documentclass[fleqn,usenatbib]{mnras}

\usepackage{newtxtext,newtxmath}

\usepackage[T1]{fontenc}
\usepackage{ae,aecompl}

\usepackage{graphicx}	
\usepackage{amsmath}	
\usepackage{amssymb}	
\DeclareMathOperator{\diag}{diag}
\usepackage{placeins}
\usepackage{xcolor}
\usepackage{bm}

\title[Accretion histories from integrated spectroscopy]{A galaxy's accretion history unveiled from its integrated spectrum}

\author[Boecker et al.]{
Alina Boecker,$^{1}$\thanks{E-mail: boecker@mpia.de}
Ryan Leaman,$^{1}$
Glenn van de Ven, $^{2}$
Mark A. Norris, $^{3}$
\newauthor
Ted Mackereth, $^{4}$
Robert A. Crain $^{4}$
\\
$^{1}$Max-Planck Institut f\"{u}r Astronomie, K\"{o}nigstuhl 17, D-69117 Heidelberg, Germany \\
$^{2}$European Southern Observatory (ESO), Karl-Schwarschild-Str. 2, 85748 Garching bei M\"{u}nchen, Germany \\
$^{3}$Jeremiah Horrocks Institute, University of Central Lancashire, Preston, PR1 2HE, UK \\
$^{4}$Astrophysics Research Institute, Liverpool John Moores University, 146 Brownlow Hill, Liverpool, L3 5RF, UK}

\date{Accepted XXX. Received YYY; in original form ZZZ}

\pubyear{2018}

\begin{document}
\label{firstpage}
\pagerange{\pageref{firstpage}--\pageref{lastpage}}
\maketitle

\begin{abstract}
We present a new method of quantifying a galaxy's accretion history from its integrated spectrum alone. Using full spectral fitting and calibrated regularization techniques we show how we can accurately derive a galaxy's mass distribution in age-metallicity space and further separate this into stellar populations from different chemical enrichment histories. By exploiting the fact that accreted lower mass galaxies will exhibit an offset to lower metallicities at fixed age compared to the in-situ stellar population, we quantify the fraction of light that comes from past merger events, that are long since mixed in phase-space and otherwise indistinguishable. Empirical age-metallicity relations (AMRs) parameterized for different galaxy masses are used to identify the accreted stellar populations and link them back to the progenitor galaxy's stellar mass. This allows us to not only measure the host galaxy's total ex-situ mass fraction ($\mathrm{f}_{\mathrm{acc}}$), but also quantify the relative amount of accreted material deposited by satellite galaxies of different masses, i.e. the accreted satellite mass function in analogy to the subhalo mass function. Using mock spectra of simulated, present-day galaxies from the EAGLE suite we demonstrate that our method can recover the total accreted fraction to within $\approx 12$ \%, the stellar mass of the most massive accreted subhalo to within $\approx 26$ \% and the slope of the accreted satellite mass function to within $\approx 16$ \% of the true values from the EAGLE merger trees. Future application of this method to observations could potentially provide us accretion histories of hundreds of individual galaxies, for which deep integrated light spectroscopy is available.
\end{abstract}




\section{Introduction}

A galaxy's total stellar mass budget can be split into two categories: \emph{in-situ} star formation that originates from cooling of the galaxy's native gas content and acquisition of \emph{ex-situ} stars through merging with satellite galaxies. The importance of the ex-situ or accreted material relative to in-situ star formation is expected to be very stochastic, but also strongly mass dependent. For example, cosmological simulations predict that higher mass galaxies with $\mathrm{M}_{\star}\gtrsim 10^{11}\ \mathrm{M}_{\odot}$ have higher ex-situ fractions between 50  and 90 \%, while Milky Way (MW) mass haloes typically only exhibit ex-situ mass fractions of 10-20 \% \citep{cooper13,cooper15,rodriguesgomez16,qu17,pillepich17}. \par 
The measurement of the ex-situ contribution to galaxy evolution is crucial to understand how mergers impact the dynamical, structural and, as we will show here, element abundances of a galaxy population. While the mass distribution of surviving and merged subhaloes is a robust prediction of $\Lambda$CDM in the absence of baryons, there exists an intrinsic degeneracy between dark and baryonic accretion histories due the effect of feedback processes on the baryons. Quantifying past accretion events in observed galaxies is therefore important, but intrinsically difficult for several reasons. For one, the remnant signatures of these events are located in observationally challenging regimes, as ex-situ dominated regions are mainly in the outskirts (> 30 kpc) of galaxies, where the surface brightness becomes very low ($\mu_g > 25 \ \mathrm{mag}/\mathrm{arcsec}^2$). Secondly, the dissolution of the merger in the host potential can smoothly distribute stellar material over a large fraction of the virial radius, which after several Gyrs, becomes difficult to identify as projected overdensities or kinematically coherent structures in external galaxies. \par 
For these reasons, the most accurate constraints on accretion histories come from deep photometric surveys of the stellar haloes of Local Group galaxies (e.g. MW: \citealp{bell08,carollo10,iorio18} and M31 \citealp{pandas,courteau11}) and beyond (e.g. \citealp{ghosts1,ghosts2,cena}), which resolve the stellar haloes into individual stars. The resolved photometry can provide information on discrete stellar populations of ex-situ stars in the diffuse outer haloes of galaxies. For example, the observed relation between the metallicity and stellar mass of the halo \citep{harmsen17} can be used to understand the mass assembly history by comparing to correlations found from cosmological simulations  \citep{font11,souzabell18,monachesi18}. \par
Other studies such as \citet{tonini13,leaman13a,kruijssen18,beasley18,mackey18,hughes18} use globular clusters (GCs) as bright tracers of past accretion events. By identifying groups of GCs as coherent structures in phase-space or with similar stellar population properties, they can be  quantitatively linked back to the stellar mass of the accreted host or the accretion time. \par
Beyond ($\gtrsim10$ Mpc) there exist also many deep photometric integrated light studies of stellar haloes, which uncover very low surface brightness substructures and streams \citep[e.g.][]{delgado10,spavone17,huang17,dragonfly,hood18}. In order to estimate an accreted fraction, the measured surface brightness profiles are usually structurally decomposed into multiple components \citep[i.e. an in-situ dominated central component and an accretion dominated "halo" component, see e.g.][]{huang16_2,iodice16,spavone17} . \par
These observational methods are restricted in the sense that they provide a single value for the ex-situ contribution or estimate bulk properties of the stellar halo. They also make structural assumptions about where the in-situ and ex-situ contributions should dominate the surface brightness spatially. However, cosmological simulations show that massive galaxies ($\mathrm{M}_{\star}\gtrsim 10^{11} \ \mathrm{M}_{\odot}$) can still have up to 60\% of accreted material in the innermost 10 kpc from the galaxy's centre \citep{pillepich17}. Furthermore, the age-metallicity degeneracy, which is especially prominent in colors from integrated photometry, means that the chemical properties are difficult to uniquely interpret. \par
Given these observational challenges, and the expected variety of merger histories in the galaxy populations, there is a clear need for an observational method that measures the accretion history of galaxies in more detail - including in their central regions as well as in a fashion that can be compared directly with simulations. If such a method can probe the luminous distribution of long disrupted substructures, which are otherwise indistinguishable, it would represent a novel test of galaxy formation models. In addition, if we want to understand the impact of the merger history on other galaxy properties like their morphology or dynamics, we need to go beyond the local universe, as only there we will acquire the needed statistics. \par
In this paper we present a novel technique to quantify a galaxy's accretion history from its integrated spectrum, and verify the method with the help of the hydrodynamical cosmological EAGLE simulations \citep{schaye15,crain15}. In section \ref{sec 2} we introduce the process of constructing the merger histories and mock spectra from the simulations. In section \ref{sec 3} we describe our method. We first model the age and metallicity \emph{distribution} of stellar populations in a galaxy with the help of full spectral fitting (pPXF: \citealp{ppxf04,ppxf17}) using updated and carefully calibrated regularization methods. The ex-situ contributions to the total stellar mass in age-metallicity space are then linked to satellite galaxy masses by empirically derived mass-dependent age-metallicity relations (AMRs). In section \ref{sec 4} we show that this method provides an observational estimate of the abundance of accreted satellite galaxies as a function of their stellar mass prior to accretion - an analogue to the unevolved\footnote{Unevolved denotes that the mass is measured before infall meaning that no mass loss due to tidal interactions has occurred yet. } subhalo mass function \citep{giocoli08,jiang16}. In section \ref{sec 5} we end with a conclusion and an outlook of our method to future applications.

\section{Simulated Spectra and Merger Histories from the EAGLE Simulation}\label{sec 2}

\begin{figure*}
\centering
	\includegraphics[width=\textwidth]{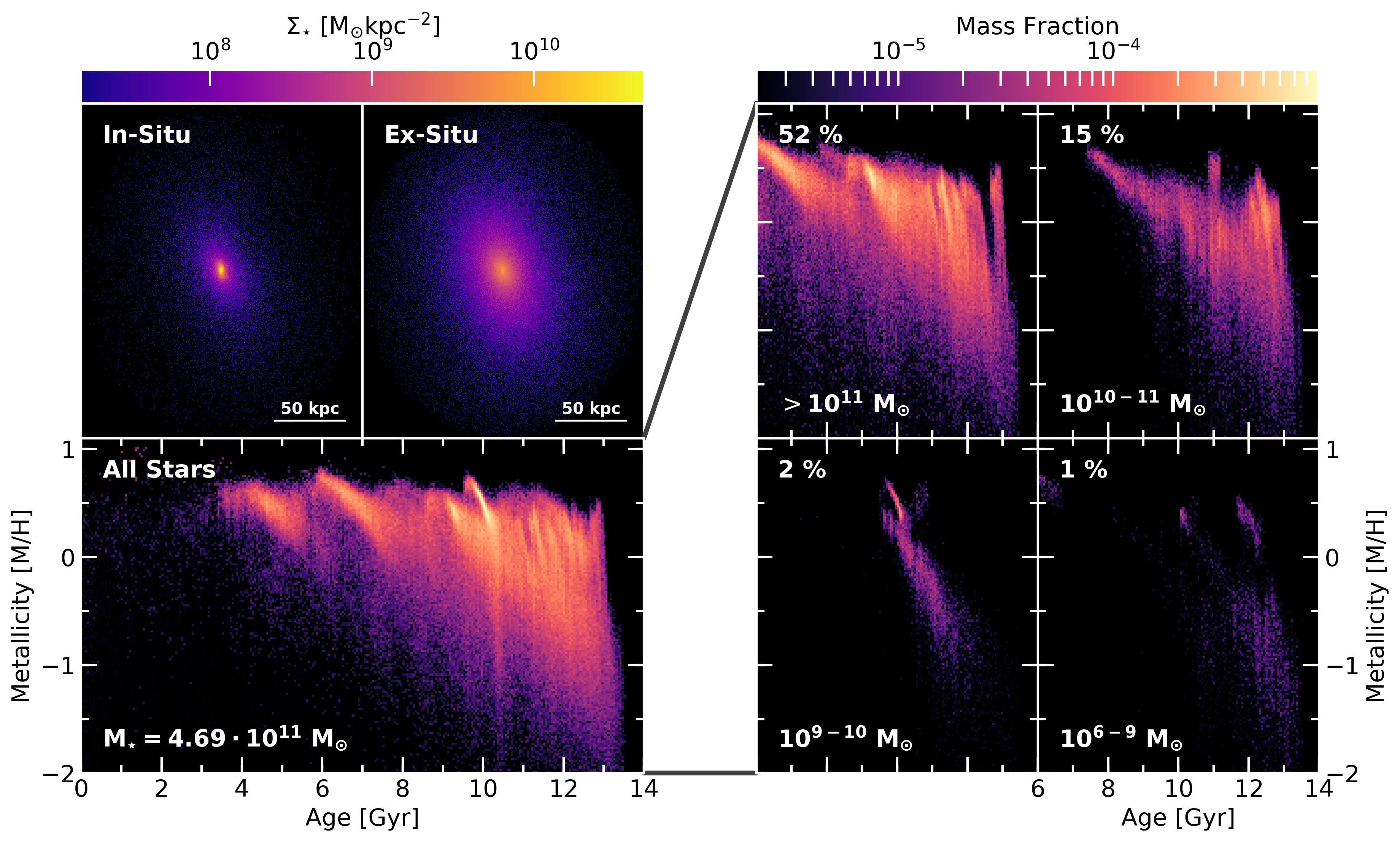}
    \caption{\textit{Upper left}: Stellar surface mass density $\Sigma_{\star}$ separated in in-situ and ex-situ particles. \textit{Lower left}: The distribution of mass fractions in age-metallicity space for all the stars of an EAGLE simulated galaxy with total stellar mass of $\mathrm{M}_{\star}=4.69\cdot 10^{11} \ \mathrm{M}_{\odot}$ (Galaxy Nr. 2, see Table \ref{tab: table1} for basic properties). \textit{Right}: The ex-situ mass distribution in age-metallicity space is further divided into contributions from progenitor satellite galaxies of different masses. The percentage numbers show the fraction of stellar mass in each mass range compared to the total stellar mass. Evidently, a single 1:1 merger brought in the majority of ex-situ stars.}
    \label{fig: plot1}
\end{figure*}

In order to test whether our new method recovers a known accretion history, we make use of galaxies from the EAGLE simulation suite \citep{schaye15, crain15}, where we have an exact knowledge of a galaxy's true mass distribution in age-metallicity space, and its accretion history. The EAGLE simulation is a hydrodynamical cosmological simulation and therefore treats the evolution of baryonic and dark matter self-consistently and adopts a series of subgrid models in order to account for baryonic physics below the resolution scale. In particular, the parameters of the subgrid models were calibrated to ensure reproduction of the $\mathrm{z}=0$ galaxy stellar mass function. The data from the EAGLE simulation have been publicly released in \citet{eagledatabase,eagleparticle}. \par
We utilize the particle information of nine early-type galaxies at $\mathrm{z}=0$ spanning a stellar mass range from $10^9$ to $10^{12} \ \mathrm{M}_{\odot}$ contained in the largest volume simulation ("REFL0100N1504"), which has a baryonic particle mass resolution of $1.81\cdot 10^6 \ \mathrm{M}_{\odot}$. This sample is not representative and was not picked for any other reason than to explore the method's success with a range of stellar masses and star formation histories. Therefore, the demonstration of our developed method throughout section \ref{sec 3} and \ref{sec 4} will focus on a single, central galaxy with a total stellar mass of $\mathrm{M}_{\star}\approx 6.77\cdot 10^{11} \ \mathrm{M}_{\odot}$, which we will refer to as Galaxy Nr. 1 according to Table \ref{tab: table1}. Additional information about basic properties of our full galaxy sample can be found in the appendix \ref{appendix}. \par
For every stellar particle we have the basic information like its mass, age, total metallicity [M/H] and position, but also the \texttt{SnapNum} (or redshift), the \texttt{GroupNumber} and \texttt{SubGroupNumber}\footnote{The \texttt{GroupNumber} and \texttt{SubGroupNumber} are integer identifiers of the Friends-of-Friends halo and subhalo respectively, which a given galaxy currently resides in. Those values are not unique across snapshots.} at which the particle first appeared as well as a flag, whether the particle was born in-situ or ex-situ. This additional information is obtained by stepping through all the snapshots and identifying particles that later end up in our selected galaxy at $\mathrm{z}=0$. Ex-situ particles are identified as those not born on the main branch of the simulated galaxy \citep{qu17}. For this study, we will focus on particles within a 3D aperture of 100 kpc with respect to the galaxy's centre of mass. \par
As we want to go one step further and associate each ex-situ particle to a host galaxy of mass $\mathrm{M}_{\mathrm{sat}}$ prior to merging with the primary halo, we step through the merger trees and track those particles. The birth \texttt{SnapNum}, \texttt{GroupNumber} and \texttt{SubGroupNumber} of every ex-situ particle in our simulated galaxy uniquely identifies the \texttt{GalaxyID}\footnote{This is a unique integer identifier of a galaxy in the simulation.} of the progenitor. By traversing up the merger tree, we can identify when the progenitor (i.e. the galaxy containing the ex-situ stars) merged onto the main branch of the host galaxy. We follow \citet{rodriguesgomez15,rodriguesgomez16} and quantify the stellar mass of the subhalo at the time $\mathrm{t}_\mathrm{max}$, where it reaches its maximum mass prior to merging - as this mass should provide the most representative indication of its chemical enrichment efficiency. \par
In the left panels of Figure \ref{fig: plot1}  we show the stellar surface mass density in the xy-plane separated into the in-situ and ex-situ components as well as the mass distribution in age-metallicity space for the entire stellar population of a EAGLE simulated galaxy with $\mathrm{M}_{\star}\approx 4.69\cdot 10^{11} \ \mathrm{M}_{\odot}$ (Galaxy Nr. 2, see Table \ref{tab: table1} for basic properties). While the ex-situ component is more spatially extended than the in-situ component, there is significant overlap spatially, and the bulk of the ex-situ component is centrally concentrated and exhibits a smooth distribution. In the right panels of Figure \ref{fig: plot1} we plot the ex-situ age-metallicity distribution further split up into the corresponding stellar mass ranges of the satellite galaxies they came from. This high mass galaxy had a 1:1 merger, but also several lower mass ratio accretion events, which brought in stellar material that is on average more metal poor at fixed age. \par
In order to test whether our machinery can successfully extract a galaxy's accretion history from its integrated spectrum alone, we construct a mock spectrum from the simulated galaxy. We represent each stellar particle within those 100 kpc as a single stellar population (SSP). We use the SSP model library MILES \citep{vaz10,vaz15} based on the BaSTI isochrones \citep{basti04, basti06}. These cover an irregular grid with 53 ages spanning from 30 Myr to 14 Gyr and 12 metallicities from -2.27 dex to 0.4 dex. We choose a bimodal initial mass function (IMF) with a slope of 1.3\footnote{The EAGLE simulation uses a Chabrier IMF \citep{chabrier03}, however as we will use the same SSP models for fitting the mock spectrum in section \ref{sec: ppxf}, this does not influence our results in any way.}  \citep{milesimfs} and added no additional dust. In this way every particle is assigned a representative spectrum suitable for its age and metallicity. The integrated spectrum of the simulated galaxy is then obtained by summing up all the SSP spectra weighted by their corresponding particle stellar mass, because the luminosity of the SSP models is expressed per unit solar mass. Naturally, the ages and metallicities of the particles in the simulation have more values than the SSP library. However, we deemed interpolating among the models was unnecessary, as the mean absolute difference (MAD) between the integrated spectrum using interpolated SSPs and non-interpolated ones was less than 0.08\%. Finally, we added random Gaussian noise to the integrated spectrum to achieve a SNR of 100 \AA$^{-1}$ for the test presented here.\footnote{In \citet{master} we investigated the SNR variations, which had an impact on the recovered distribution in age-metallicity space from regularized full spectral fitting, but was less significant with regard to the accreted satellite mass function.}

\section{Methodology}\label{sec 3}

\subsection{Recovering an extended mass distribution in age-metallicity space from an integrated spectrum}\label{sec: ppxf}

\begin{figure*}
\centering
	\includegraphics[width=\textwidth]{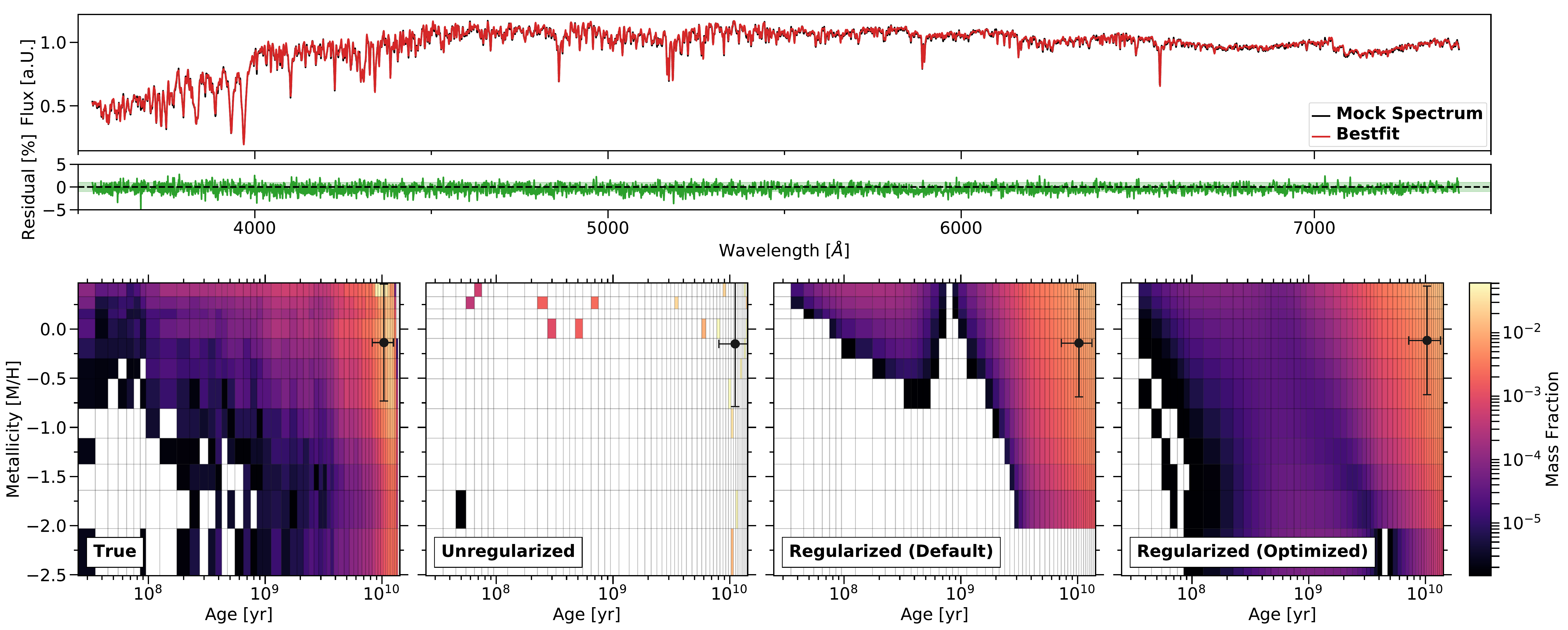}
    \caption{\textit{Top}: Comparison between the mock spectrum (\textit{black}) and the best-fit (\textit{red}) from calibrated regularized least squares with the third order difference operator as a regularization matrix. The residuals (\textit{green}) are on the order of 1 \%, which is expected for a SNR of 100. This plot looks identical to the eye for the unregularized and default regularized case. \textit{Bottom}: Mass distribution in age-metallicity space for the EAGLE simulated galaxy (Galaxy Nr. 1). Color coding corresponds to the mass fraction in each age-metallicity bin and is identical for all four panels. The errorbar shows the mass-weighted mean and standard deviation of age and metallicity. \textit{From left to right}: True distribution of mass fractions in the EAGLE galaxy binned to the available age-metallicity grid of the MILES SSP models; recovered mass fractions from full spectral fitting using normal linear least squares (no regularization); calibrated regularized least squares with the default regularization matrix (second order difference operator); and with the optimized regularization matrix (third order difference operator).}
    \label{fig: plot2}
\end{figure*}

The first step in our method is to use full spectral fitting, where the mock integrated galaxy spectrum is fitted with a linear combination of SSP model templates in order to infer stellar population parameters like age and metallicity. We do this by using pPXF \citep{ppxf04, ppxf17}, which optimizes the weights for the linear combination of  SSP templates and hence provides us with a \emph{distribution} in age-metallicity space instead of average quantities. Because SSP models are normalized to one solar mass, the recovered weights are mass fractions. For simplicity and to better understand the sources of uncertainty in our final quantity of interest, we fit the same SSP models that were also used when constructing the mock spectrum. \par
This inverse problem can in principle be solved by a simple linear least squares minimization. However, if we compare the solution from this with the true distribution of the EAGLE particle data for our particular galaxy binned to the SSP model grid in Figure \ref{fig: plot2}, we see that the recovered mass fractions are distributed very sparsely and not necessarily in a physically meaningful way across the age-metallicity plane. This behaviour is due to the nature of the SSP models, as their shapes are determined by stellar physics and evolution models and are hence degenerate. Therefore, the optimal representation of the mock spectrum can be found with an ambiguous and almost arbitrary combination of SSP templates (ill-posed problem). Furthermore, due to the flexibility of optimal weight combinations, any change in the input galaxy spectrum due to noise can change the solution drastically (ill-conditioned problem). \par 
pPXF offers a way to circumvent those problems via regularization. Mathematically speaking this dampens weight solutions that are driven by noise in the data. In an astrophysical context, it ensures a certain smoothness between neighbouring SSP bins in age-metallicity space, which is appropriate as chemical enrichment proceeds smoothly in galaxies as seen in Figure \ref{fig: plot1}. The amount of smoothness in pPXF is controlled by the \emph{regularization parameter} $\lambda$ and the way the weights are smoothed out is defined by the \emph{regularization matrix} $\bm{B}$. \par
Crucially, $\lambda$ needs to be calibrated for \emph{every} spectrum that is fit, as too low or too high regularization parameters will recover an astrophysically different star formation history and chemical enrichment. We follow the calibration procedure that is described in the source code of pPXF, which was adopted from \citet[][section 19.4.1]{press07} and has been used in astronomical literature \citep[see e.g.][]{norris15,mcdermid15}. The "optimal" regularization parameter represents the maximal amount of smoothness for a given data fidelity, such that the solution is still consistent with the unregularized solution. This is found in the following way: first the noise vector is re-scaled such that the reduced $\chi^2$ of the unregularized fit becomes unity. Then a series of fits adopting different regularization parameters is performed until the $\chi^2$ has increased by one standard deviation, which corresponds to a $\Delta\chi^2$ of $\sqrt{2N}$, where $N$ is the number of pixels included in the fit. \par
The default regularization matrix that is adopted in pPXF is the second order finite difference operator, $\bm{B}=\diag(1,-2,1)$, while the user also has the option to choose the first order one, $\bm{B}=\diag(1,-1)$. However, we have found that neither of those two options recover the very extended distribution in age-metallicity for our given simulated galaxies well enough. We therefore manually introduced a third order finite difference operator, $\bm{B}=\diag(1,-3,3,-1)$, in the source code of pPXF, which seems to recover the overall shape of the true distribution much better. The optimal regularized solution following the calibration, and using the default as well as our implemented matrix is shown in Figure \ref{fig: plot2} for comparison. The median absolute deviation (MAD) of the absolute residuals between the true mass distribution in age-metallicity space and the recovered one with the updated regularization matrix is 14 \% lower than for the default matrix. \par
We also show in Figure \ref{fig: plot2} the best-fit spectrum for the regularized solution using the third order difference operator, however we note that in practice this is indistinguishable from the unregularized and default regularized case. The standard deviation of the residuals are 1 \%, which shows that the spectrum is fitted down to the injected noise level with SNR of 100 even for the regularized case.

\subsection{Flexible mass-dependent chemical enrichment templates} 

With a robust recovery of extended mass distributions in age-metallicity space from regularized pPXF fitting, we now need to associate the mass fractions to potential accreted galaxies. We do this by constructing flexible, mass-dependent templates in age-metallicity space, which describe how galaxies of a given mass should (on average) chemically evolve. While the detailed chemical evolution of distant high mass galaxies is not observationally constrained, here we attempt to construct a physically motivated, flexible, mass-dependent chemical framework. \par
For this we use results from \citet{leaman13a}, who derived empirical age-metallicity relations spectroscopically for local group dwarf galaxies from resolved stellar populations. To first order, leaky box chemical evolution models describe the metallicity distribution function (MDF) and age-metallicity relation (AMR) of those galaxies with only a galaxy mass-dependent variation in the effective yield $p(\mathrm{M}_{\star})$. For a given galaxy, the chemical evolution is described as:
\begin{equation}\label{eq: 1}
Z(t)=-p(\mathrm{M}_{\star})\ln \mu (t)\,
\end{equation}
where $Z$ is the metallicity, $t$ (in Gyr) is time since the Big Bang, $\mu$ the galaxy's gas fraction and $p(\mathrm{M}_{\star})$ the mass-dependent effective yield. \par
The mass-dependent effective yield is empirically measured for the Local Group galaxies below $\mathrm{M}_{\star}\leq 10^{9} \ \mathrm{M}_{\odot}$ in \citet{leaman13a} \citep[see also][]{lee06} and yields the observed relation of $p(\mathrm{M}_{\star}) \propto \mathrm{M}_{\star}^{\alpha_p}$ with $\alpha_p\simeq 0.4$. Above this mass, the observed mass-metallicity relation (MMR) is seen to flatten \citep[e.g.][]{gallazzi05}, and we take this into account by modifying the functional form of the $p(\mathrm{M}_{\star})$-relation such that a galaxy's average stellar mass will reproduce the turnover in the MMR (at $\sim 10^{10} \ \mathrm{M}_{\odot}$), while still matching the MDF of low mass galaxies:
\begin{equation}\label{eq: 2}
\log_{10}p(\mathrm{M}_{\star})=p_0 +\log_{10}\left(1-\exp{\left(-\frac{\mathrm{M}_{\star}}{\mathrm{M}_0}\right)}^{\alpha_p}\right),
\end{equation}
here $p_0$ describes the value the relation asymptotes towards for high galaxies masses, $\mathrm{M}_0$ is the turn-over mass and $\alpha_p$ is the low-mass slope. \par
In order to convert the iron abundances [Fe/H], as they were derived in \citet{leaman13a}, to total metallicity [M/H], which is used in the SSP model grid, we adopt the following relation from \citet{sc05} 
\begin{equation}
[\mathrm{M}/\mathrm{H}]=[\mathrm{Fe}/\mathrm{H}]+\log_{10}\left(0.694\cdot 10^{[\alpha/\mathrm{Fe}]}+0.306\right).
\end{equation}
We further allow each galaxy to have a mass-dependent evolution in [$\alpha$/Fe], by utilizing the empirical [$\alpha$/Fe]-[Fe/H]-relation derived from individual stars on Local Group galaxies in \citet{deboer14}. They found that the "knee" in the [$\alpha$/Fe]-[Fe/H] diagram ([Fe/H]$_{\mathrm{knee}}$) occurs at higher metallicity for high mass galaxies \citep[see also][]{kirby11}. We adopt  this mass dependence of the [Fe/H]$_{\mathrm{knee}}$ position as well as of the slope of the low alpha sequence. We calibrate the [$\alpha$/Fe]-plateau value to have a mass dependence as well, such that the mean $\alpha$-abundance versus galaxy mass trends seen in observations \citep{thomas10} and EAGLE \citep{segers16} are reproduced. \par
Lastly, we account for a mass-dependent gas fraction evolution $\mu (t)$, which can be due to gas being consumed in star formation and/or being removed through feedback processes or quenching. Higher mass galaxies ($\gtrsim 10^{10} \ \mathrm{M}_{\star}$) typically exhaust their in-situ gas much quicker than lower mass galaxies \citep[e.g.][]{mcdermid15,pacifici16a,pacifici16b} and we therefore parameterize the gas fraction as:
\begin{equation}\label{eq: 3}
\mu(t)=\frac{t-\left(13.5-t_{\mathrm{form}}\right)}{t_{\mathrm{form}}},
\end{equation}
where $t_{\mathrm{form}}$ is an epoch by which the galaxy has formed its in-situ stars. We allow a galaxy mass dependence to enter through this formation time as:
\begin{equation}
t_{\mathrm{form}}=\min\left[\left(14-\log_{10}\ \mathrm{M}_{\star}\right)^{\alpha_t},\ 13.5\right]
\end{equation}
with $\alpha_t$ influencing how long the star formation duration will be for a given galaxy mass. This will result in AMR curves that do not evolve until $\mathrm{z}=0$, but reach their maximum metallicity at $t_{\mathrm{form}}$ for higher galaxy masses ($\gtrsim 10^{10} \ \mathrm{M}_{\star}$). \par
A set of these AMR templates for different galaxy masses and with parameters $p_0=0.1$, $M_0=10^{10.5} \ \mathrm{M}_{\odot}$, $\alpha_p=0.4$ and $\alpha_t=2.0$ is plotted in Figure \ref{fig: plot3}a) as an example. Importantly however, these parameterizations enable us to flexibly vary the shape of the mass dependent chemical evolution tracks allowing for a stochastic assessment of the uncertainties in our final quantities of interest (see section \ref{uncertain}). 

\subsection{Associating the spectroscopic mass fractions to accreted satellite galaxies}

\begin{figure*}
\centering
	\includegraphics[width=\textwidth]{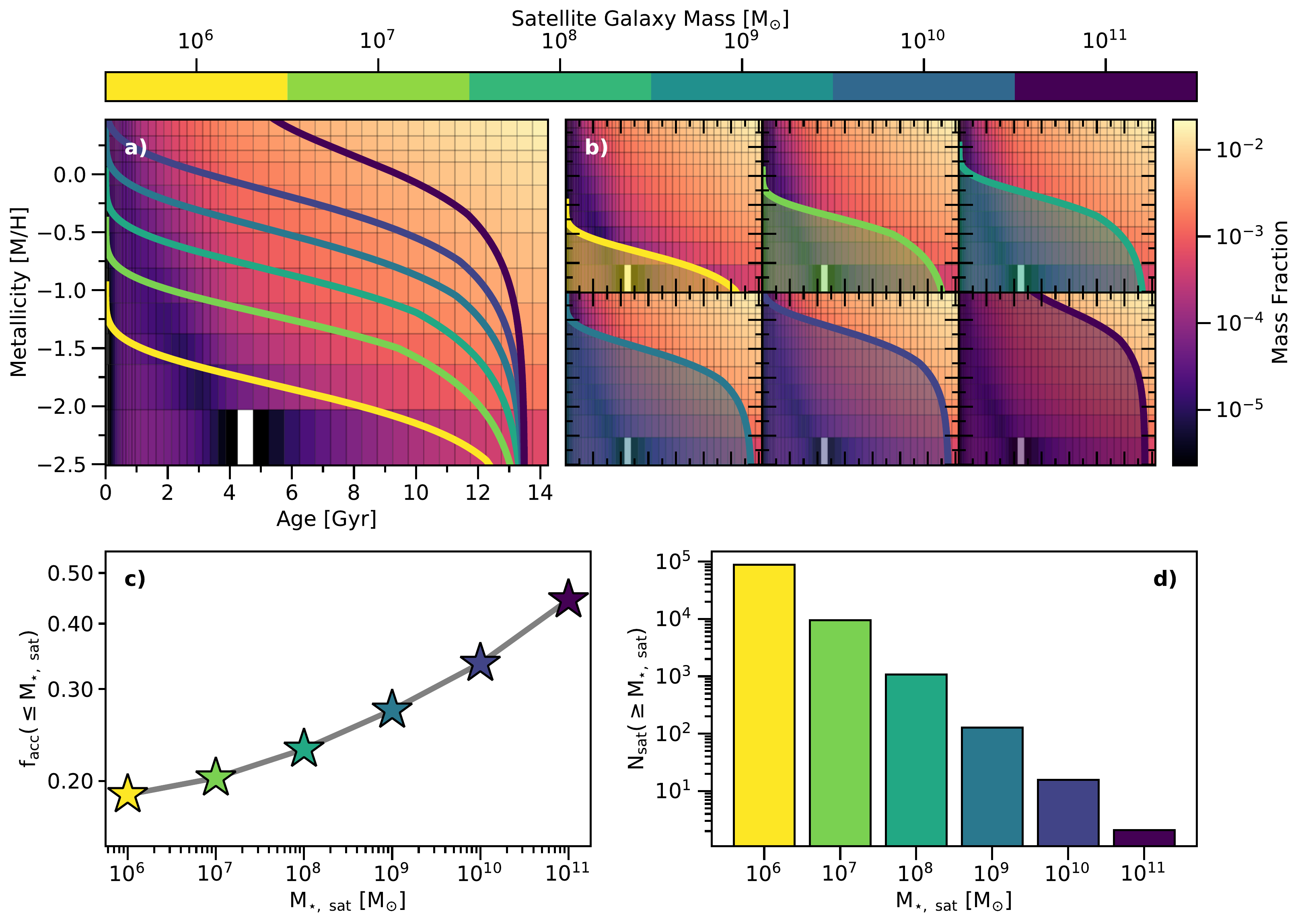}
    \caption{Schematic of our method. This shows just one realization of the chemical evolution templates. \textit{a)}: Mass-dependent age-metallicity relation templates for galaxy masses in the range of $10^6 \ \mathrm{M}_{\odot}$ and $10^{11} \ \mathrm{M}_{\odot}$ (see colorbar) overplotted onto the pPXF recovered mass fractions in age-metallicity space for Galaxy Nr. 1 as per Figure \ref{fig: plot2}. \textit{b)}: Visualization of how the accreted satellite mass function is constructed. Mass weights coinciding in the shaded regions, i.e. below each age-metallicity relation curve are summed up. This then represents the cumulative accretion fraction brought in by a satellite galaxy with a stellar mass corresponding to the associated mass-dependent age-metallicity relation. \textit{c)}: Resulting cumulative accretion fractions. The color of the star symbols correspond to the respective galaxy mass of the age-metallicity template from which the accretion fraction was calculated. \textit{d)}: An analogue to the accreted satellite mass function calculated from the accretions fractions found with our method.}
    \label{fig: plot3}
\end{figure*}

With the galaxy mass-dependent chemical templates described above, we are able to link the mass fractions in age-metallicity space recovered from the regularized pPXF solutions to astrophysical quantities of interest - such as the galaxy's total fraction of accreted mass as well as the distribution of merged satellite galaxies. The amount of mass in accreted satellites of a given mass can be straightforwardly computed by overlaying our mass-dependent age-metallicity relation templates onto the spectroscopically recovered mass distribution in age-metallicity space, as seen in Figure \ref{fig: plot3}a). Every mass weight $m_i$ recovered from pPXF lying \emph{below} a certain AMR curve is treated as potentially coming from accreted satellite galaxies with chemical evolution representative of that mass or lower, i.e. $\mathrm{f}_{\mathrm{acc}}(\leq \mathrm{M}_{\mathrm{sat}})=\sum_i m_i(t, [\mathrm{M}/\mathrm{H}]\leq [\mathrm{M}/\mathrm{H}]_{\mathrm{template}}(t | \mathrm{M}_{\ \mathrm{sat}}))$. \par
This method of recovering the contributions from accreted satellite galaxies of a given mass is only considered valid in a cumulative sense, as there is an astrophysical degeneracy inherent to the assignment of mass fractions of low-mass satellites. We do not know a priori, whether these recovered stellar populations with lowest [M/H] at fixed age are coming from low-mass satellites directly accreted to the host or if they were first accreted to an intermediate mass satellite, which then merged with the host. \par 
The chemical evolution templates extend up to an AMR associated with some most massive accreted satellite galaxy, $\mathrm{M}_{\mathrm{sat, \ max}}$. At this mass, the method has provided an estimate of the total accreted fraction for the host galaxy, i.e. $\mathrm{f}_{\mathrm{acc}, \ \mathrm{tot}}=\mathrm{f}_{\mathrm{acc}}(\leq \mathrm{M}_{\mathrm{sat, \ max}})$. Every mass fraction lying above the AMR template associated with $\mathrm{M}_{\mathrm{sat, \ max}}$ is considered \emph{in-situ} according to our method, but naturally there will be some overlap with the ex-situ contributions, as the age and metallicity properties are very similar in that mass regime. \par
In Figure \ref{fig: plot3} we show a schematic of our method to associate the spectroscopically recovered mass fractions with accreted satellite galaxies of different masses. AMR templates corresponding to accreted galaxies of masses between $10^6$ and $10^{11} \ \mathrm{M}_{\odot}$ are overlaid on the spectroscopically recovered mass fractions in Figure \ref{fig: plot3}a). The mass fractions below an AMR curve represents the contribution to the galaxy's merger history from satellite galaxies of this mass as seen in Figure \ref{fig: plot3}b). The resultant cumulative accretion fractions versus the associated satellite galaxy masses are then produced by summing up the mass fractions lying in the shaded regions respectively and are plotted in Figure \ref{fig: plot3}c). In Figure \ref{fig: plot3}d) we show an analogue to the (unevolved) subhalo mass function, which can be calculated by dividing the recovered cumulative accreted mass by its associated satellite galaxy mass (i.e. $\mathrm{N}_{\mathrm{sat}}(\geq \mathrm{M}_{\mathrm{sat}})=\mathrm{M}_{\mathrm{acc}}(\leq \mathrm{M}_{\mathrm{sat}})/\mathrm{M}_{\mathrm{sat}}$, where $\mathrm{M}_{\mathrm{acc}}(\leq \mathrm{M}_{\mathrm{sat}})=\mathrm{f}_{\mathrm{acc}}(\leq \mathrm{M}_{\mathrm{sat}})\cdot \mathrm{M}_{\mathrm{host}}$). \par
Figure \ref{fig: plot3} shows the recovered accreted mass fractions for \emph{one} realization of the chemical evolution templates ($p_0=0.1$, $M_0=10^{10.5} \ \mathrm{M}_{\odot}$, $\alpha_p=0.4$ and $\alpha_t=2.0$), however in the final results (section \ref{sec 4}) we show them as the median of many realizations (see section \ref{uncertain} for more details).

\subsubsection{Stochastically assessing the systematic uncertainties}\label{uncertain}

As the shape of the chemical evolution templates is not well constrained by observations, especially for galaxy masses higher than found in the Local Group, we introduce flexibility and perform a Monte-Carlo (MC) simulation. We randomly draw for 1000 trials the parameters $p_0$, $\mathrm{M}_0$, $\alpha_p$ (equation \ref{eq: 2}) and $\alpha_t$ (equation \ref{eq: 3}) from a uniform distribution in a range of [-0.2,0.2], $10^{[9.5,11]}$, [0.4,0.6] and [1.7,2.5] respectively. We also add a scatter to the sampled AMR curve, which is drawn from a Gaussian distribution with a standard deviation of $\pm 0.2 \ \mathrm{dex}$ in order to account for the intrinsic scatter of a galaxy's metallicity at fixed stellar age \citep[see e.g.][]{leaman13a}. The choices of the parameter variation ranges were made such that the calculated MMR from the median of the mass-dependent AMR curves lie in the scatter of the observed MMR \citep{gallazzi05,kirby13}. Uncertainties in the derived [$\alpha$/Fe]-[Fe/H]-mass relation are not accounted for, as their impact is negligible in comparison to the other parameter variations. Hence, the recovered cumulative accretion fraction as a function of satellite galaxy mass is the median of the 1000 MC trials and the uncertainty is expressed by the 16th and 84th percentiles of the trials.  \par
While this takes into account systematic uncertainties, random errors due to signal-to-noise variations in the integrated spectrum or the exact nature of the regularization calibration procedure are not considered. We however tested the variation in the shape of the recovered satellite mass function for SNR of 50, 100, 200 and 500 as well as the first, second and third order difference operator as the regularization matrix. The scatter was found to be much smaller than the uncertainties of the chemical evolution templates.

\subsubsection{Defining the most massive accretion event}

In principle, the cumulative accretion fraction corresponding to the highest satellite galaxy mass marks the  \emph{total} accretion fraction, $\mathrm{f}_{\mathrm{acc}, \ \mathrm{tot}}$, of the host galaxy. However, as we do not have any prior knowledge about the most massive satellite ever accreted by the host, $\mathrm{M}_{\mathrm{sat, \ max}}$, our derived curve extends arbitrarily higher than the true total accretion fraction. \par
We approximate the "stopping point" by calculating the stellar mass of the accreted satellite galaxy, where the number of accreted satellite galaxies, $\mathrm{N}_{\mathrm{sat}}$, becomes unity on our spectroscopically recovered accreted satellite mass function. In return, we can truncate the cumulative accretion fractions at that point and derive the total accretion fraction of the host galaxy. Similarly, errors are computed by the intersection of the 16th and 84th percentile uncertainty of the accreted satellite mass function with unity.\par
This procedure works quite well even though our method can only compute a lower limit of the accreted satellite mass function, as it does not take into account any higher orders of subhalo-subhalo mergers. We found that this gives better results than using theoretical predictions of cosmological simulations, which can provide us with a statistical relation of the stellar mass of the most massive accreted satellite for a given host galaxy \citep[see e.g.][]{souzabell18}. In addition, it allows us to implicitly characterize $\mathrm{M}_{\mathrm{sat, \ max}}$ as an intrinsic measure of our method.

\section{Results}\label{sec 4}

\begin{figure*}
\centering
	\includegraphics[width=\textwidth]{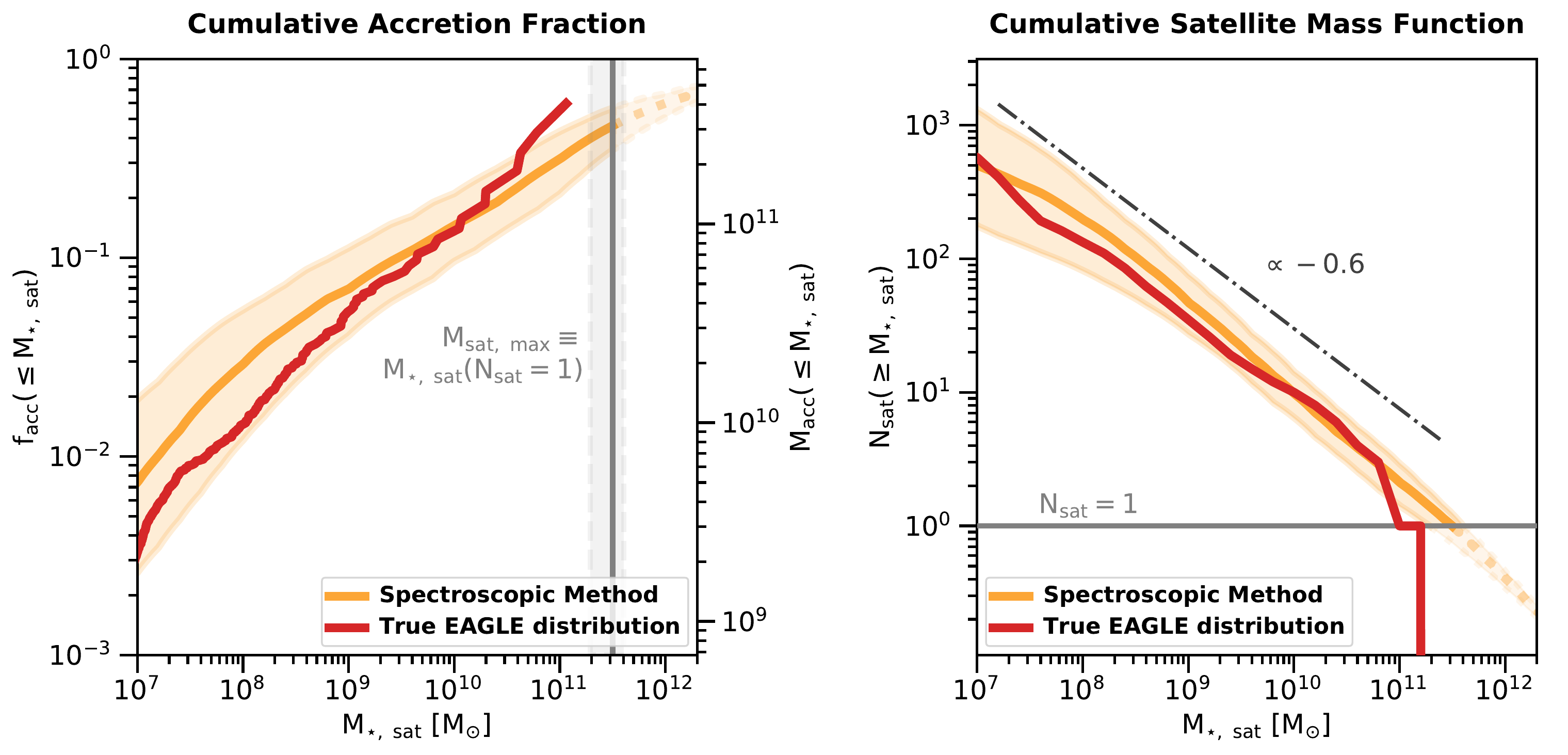}
	\caption{\textit{Left}: Cumulative accretion fraction (accreted mass) versus the stellar mass of the accreted satellite galaxy for Galaxy Nr. 1. The red line shows the true function from the simulation, while the orange line shows the function recovered from a single integrated spectrum with our method (median of the Monte-Carlo trials). The orange band marks the 16th and 84th percentile of those trials. The most massive accreted satellite ($\mathrm{M}_{\mathrm{sat, \ max}}$), which marks the total accreted fraction, is shown by the grey vertical line. \textit{Right}: The cumulative number of accreted satellite galaxies versus their stellar mass for Galaxy Nr. 1. The stopping point as shown by the light grey horizontal line, where $\mathrm{N}_{\mathrm{sat}}=1$, determines the most massive accreted satellite ($\mathrm{M}_{\mathrm{sat, \ max}}$). The approximate slope of the accreted satellite mass function is represented by the dark grey dashed-dotted line and is $\sim -0.6$.}
	\label{fig: plot4}
\end{figure*}

Having optimized the recovery of extended age-metallicity distributions, and formulated a way to link these to accreted galaxies of different masses, we can proceed with an example on a massive EAGLE galaxy (Galaxy Nr. 1) - one where we can independently verify the spectroscopically recovered accretion history by comparing to the known merger tree. \par
In the left panel of Figure \ref{fig: plot4} we show the derived cumulative accretion fraction as a function of accreted satellite galaxy mass as well as the uncertainties as calculated from the section above. We computed the cumulative accretion fraction with AMRs corresponding to galaxy masses between $10^{6.2}$ and $10^{12.4} \ \mathrm{M}_{\odot}$ in 0.2 dex steps. \par
The agreement with the true accretion fractions obtained from the EAGLE merger trees is remarkable given that we obtained the "observed" quantity purely from a simulated integrated spectrum. This result suggests that the assumptions behind the mass-dependent chemical evolution templates are reasonable and provide a novel way to recover signatures of otherwise unobservable ancient merger events. For this example galaxy, we recover a total accreted fraction of $\mathrm{f}_{\mathrm{acc}, \ \mathrm{tot}}=0.461^{+0.104}_{-0.113}$, while the "true" total accreted fraction from the ex-situ particle classification is 0.421. For the most massive progenitor we find a stellar mass of $\log_{10}\mathrm{M}_{\mathrm{sat, \ max}}=11.50^{+0.11}_{-0.21}$ dex, whereas the actual value is 11.06 dex. The errors have been calculated by the intersections of the scatter of the MC trials of the recovered satellite mass function and where $\mathrm{N}_{\mathrm{sat}}=1$.  \par
In the right panel of Figure \ref{fig: plot4} we compare the true satellite mass function of the EAGLE galaxy with our estimate, which also shows excellent agreement. We measured the slope ($\alpha_{\mathrm{sat}}$) of the recovered and true satellite mass function by fitting a power law between $10^7 \  \mathrm{M}_{\odot}$ and $\mathrm{M}_{\mathrm{sat, \ max}}$ in log space. We found the values for $\alpha_{\mathrm{sat}}$ to be $-0.65^{+0.08}_{-0.17}$ and $-0.60$ respectively. The errors where estimated by calculating the slope of the extreme points of the recovered satellite mass function by taking the scatter of the MC trials as well as the uncertainty of $\mathrm{M}_{\mathrm{sat, \ max}}$ into account. The agreement of our result with the true satellite mass function is somewhat surprising as our method is formally providing a lower limit to the satellite mass function, as it cannot differentiate mergers that happened prior to a galaxy merging to the primary halo. For example,  if a late time merger of high mass had its own accretion history, this would be degenerate with our solutions resulting in a flattening of the slope of the $\mathrm{f}_{\mathrm{acc}}$-$\mathrm{M}_{\mathrm{sat}}$ relation and steepening of the $\mathrm{N}_{\mathrm{sat}}$-$\mathrm{M}_{\mathrm{sat}}$ relation. However, we expect that this effect will be very small and well inside our uncertainties \citep[see e.g.][Figure 9]{jiang16}. Nevertheless, the validation of this method using the EAGLE simulations, suggests it is a powerful new way to recover the accreted satellite mass function and distribution of merger mass ratios in galaxies. \par
We have also performed the same analysis on eight more EAGLE simulated galaxies, which all lie in a mass range of $10^9$ to $10^{12} \ \mathrm{M}_{\odot}$. This is by no means a statistical or comprehensive sample, but we wanted to illustrate that this method works on more than one handpicked galaxy. In Figure \ref{fig: plot5} and \ref{fig: plot6} we show a comparison between the true parameters of $\mathrm{f}_{\mathrm{acc, tot}}$, $\mathrm{M}_{\mathrm{sat, \ max}}$ and $\alpha_{\mathrm{sat}}$ and those retrieved from an integrated spectrum with our method. They overall lie on the 1:1 relation within the uncertainties suggesting that this method performs well on different galaxies. The mean accuracy averaged across all nine analyzed EAGLE galaxies is 12 \%, 26 \% and 16 \% for $\mathrm{f}_{\mathrm{acc, tot}}$, $\mathrm{M}_{\mathrm{sat, \ max}}$ and $\alpha_{\mathrm{sat}}$ respectively.

\section{Conclusion and Outlook}\label{sec 5}

In this work we have presented and validated a new method of measuring a galaxy's accretion history from its integrated spectrum alone. Not only are we able to quantify the total accreted fraction of a galaxy, but also measure relative amounts of accreted material coming from merged galaxies of different masses. \par
We hence provide, for the first time, an observational method to asses the fraction of accreted material from completely disrupted satellites as a function of their original stellar mass prior to merging. This is possible, because we exploit the fact that the chemical evolution of a galaxy varies for different masses, and hence accreted material does not lie in the same region in age-metallicity space as in-situ created material. Other studies such as \citet{tonini13,leaman13a,kruijssen18,beasley18} have also successfully exploited similar arguments to quantitatively link globular clusters to the total accreted fraction of their host. Importantly, identifying these signatures from a spectrum is \emph{only} possible, if the full spectral fitting code can recover a distribution simultaneously in age and metallicity. \par
The advantage of this method is that it provides a more detailed understanding of the accretion history of a galaxy than existing methods \citep{spavone17,huang16_2,cronjevic17,harmsen17,souzabell18,monachesi18}, while simultaneously being applicable to potentially every galaxy at low or high redshift, for which an integrated spectrum exists. In particular, it is able to make predictions about completely disrupted past merger events, which are otherwise indistinguishable, because they are no longer evident in phase-space in the form of streams or shells. Because this method also links chemical signatures recovered in the spectrum back to the original stellar mass of the accreted satellite galaxy prior to merging, it is so far the closest observational equivalent to a simulation's merger tree. \par
In future work, we will perform more tests that will especially focus on comparing the recovery of the accretion history when only considering the light contribution of the centre or the outskirts of a galaxy. This will grant us insight as to whether there will be enough line-of-sight integration of ex-situ material in the halo, such that a central high signal-to-noise spectrum is sufficient for our method. If that is the case, this method could also be applied to higher redshift galaxies. Next, we will focus on observationally verifying this technique on targets, where there is an alternative measure of stellar population properties from resolved stars available. This is important in the sense that we can determine, whether this method is robust against difficult-to-model observational complications, as well as being able to fine tune the fitting technique. The necessary high signal-to-noise spectrum can observationally be achieved with today's modern integral field units such as MUSE \citep{muse}. \par 
In summary, our presented method provides a novel opportunity to validate aspects of hierarchical structure formation, as well as to make a connection to fundamental physical processes on galactic scales such as quenching or dynamical transformations. Together these will help us understand the different evolutionary pathways galaxies can experience in their mass assembly, eventually leading to the large diversity in galaxies we observe today.\par

\section*{Acknowledgements}

RL acknowledges support from the Natural Sciences and Engineering Research Council of Canada PDF award, and by Sonderforschungsbereich SFB 881 "The Milky Way System" (subproject A7and A8) of the German Research Foundation (DFG) and DAAD PPP exchange program Projekt-ID 57316058. GvdV acknowledges funding from the European Research Council (ERC) under the European Union's Horizon 2020 research and innovation programme under grant agreement No 724857 (Consolidator Grant ArcheoDyn). RAC is a Royal Society University Research Fellow. This work made use of high performance computing facilities at Liverpool John Moores University, partly funded by the Royal Society and LJMU's Faculty of Engineering and Technology.




\bibliographystyle{mnras}
\bibliography{letter} 



\appendix

\section{EAGLE Galaxy Sample}\label{appendix}

\begin{figure*}
\centering
\includegraphics[width=\textwidth]{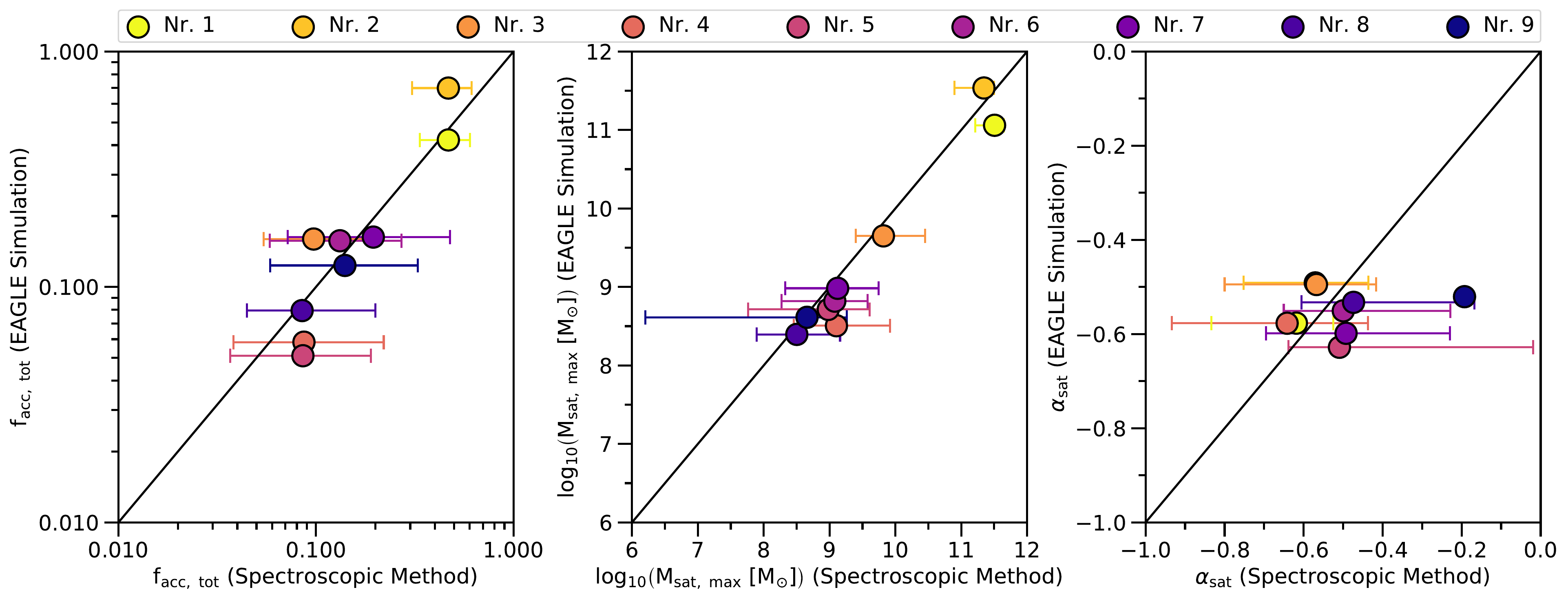}
\caption{Comparison between the total accreted fraction $\mathrm{f}_{\mathrm{acc, \ tot}}$, the stellar mass of the most massive accreted satellite galaxy $\mathrm{M}_{\mathrm{sat, \ max}}$ and the slope of the accreted satellite mass function $\alpha_{\mathrm{sat}}$ recovered with our spectroscopic method and the truth from the EAGLE simulation for all nine EAGLE galaxies that we investigated in this work. The galaxy number (Nr.) is sorted from highest (lighter color) to lowest (darker color) stellar mass. The errorbars were computed from the scatter in the MC simulations by varying the shape of the chemical evolution templates. All displayed quantities for every galaxy were computed using a SNR of 100 for the mock spectrum and the third order difference operator as the regularization matrix. The black line marks the one-to-one relation.}
\label{fig: plot5}
\end{figure*}

In total we have applied this method to nine EAGLE simulated galaxies in a mass range between $10^9$ and $10^{12} \ \mathrm{M}_{\odot}$, as summarized in Table \ref{tab: table1}. We followed the exact same procedure as described throughout this paper. In Figure \ref{fig: plot5} we compare the parameters of the satellite mass function, $\mathrm{f}_{\mathrm{acc, \ tot}}$, $\mathrm{M}_{\mathrm{sat, \ max}}$ and $\alpha_{\mathrm{sat}}$, derived with our spectroscopic method and the truth from the EAGLE merger trees. Almost all lie on the 1:1 relation. \par
\begin{figure*}
\centering
\includegraphics[width=\textwidth]{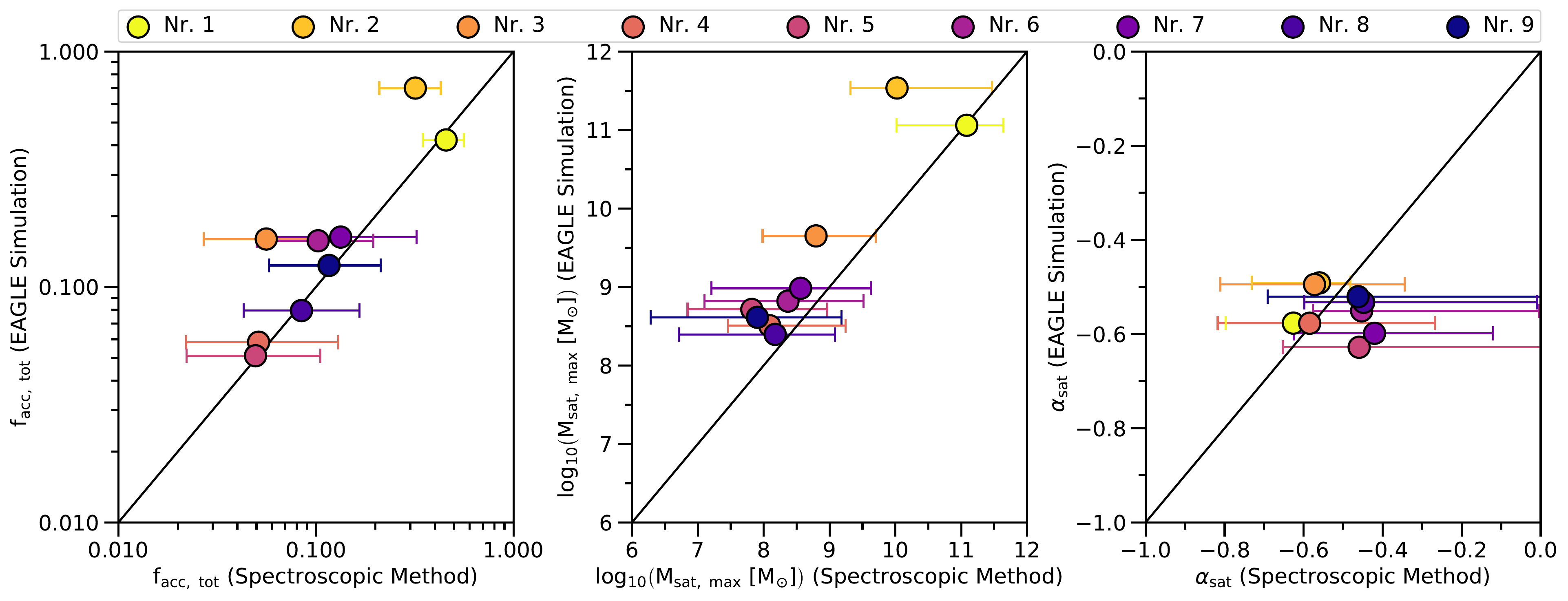}
\caption{Same as Figure \ref{fig: plot5} but now the displayed quantities for every galaxy are averages across 12 satellite mass functions measured from a mock spectrum of  SNR of 50, 100, 200 and 500 as well as the first, second and third order difference operator as the regularization matrix.}
\label{fig: plot6}
\end{figure*}

In addition, we derived the mass distribution in age-metallicity space for a mock spectrum with SNR of 50, 100, 200 and 500 as well as for the first, second and third order difference operator as a regularization matrix, which in total makes 12 derived satellite mass functions for every simulated galaxy. The averages of the derived parameters, $\mathrm{f}_{\mathrm{acc, \ tot}}$, $\mathrm{M}_{\mathrm{sat, \ max}}$ and $\alpha_{\mathrm{sat}}$, of those 12 variations for our galaxy sample are summarized in Figure \ref{fig: plot6}. Consequently, the errorbars in this plot do not only encompass the systematic uncertainties in the chemical evolution, but also random uncertainties, when deriving the age-metallicity distribution from full spectral fitting. The mean accuracy averaged across all analyzed EAGLE galaxies and the 12 realizations is 12 \%, 26 \% and 16 \% for $\mathrm{f}_{\mathrm{acc, tot}}$, $\mathrm{M}_{\mathrm{sat, \ max}}$ and $\alpha_{\mathrm{sat}}$ respectively. \par
This sample of nine EAGLE galaxies is by no means comprehensive, but it shows that this method performs well on different galaxies with different accretion histories and across a wide range of masses.

\begin{table}
\centering
\caption{Overview of the nine EAGLE simulated galaxies. \textit{From left to right column}: Nr.,  \texttt{GalaxyID}, total stellar mass inside 100 kpc as well as the mass-weighted mean age and metallicity.}
\label{tab: table1}
\begin{tabular}{lllrr}
\hline
Nr. & \texttt{GalaxyID} & $\mathrm{M_{\star}}$ & $\langle \mathrm{Age} \rangle$ & $\langle [\mathrm{M/H}] \rangle$ \\ 
& & $[\mathrm{M_{\odot}}]$  & $[\mathrm{Gyr}]$ & $ [\mathrm{dex}]$ \\
\hline
1 & 21109760 & $6.77\cdot 10^{11}$ & 10.43 & -0.068 \\
2 & 20163968 & $4.69\cdot 10^{11}$ & 9.83 & -0.052 \\
3 & 16971377 & $6.72\cdot 10^{10}$ & 9.62 & 0.135 \\
4 & 9944856 & $1.46\cdot 10^{10}$ & 8.24 & 0.050 \\
5 & 5368409 & $1.11\cdot 10^{10}$ & 9.26 & 0.008 \\
6 & 12092377 & $9.09\cdot 10^{9}$ & 9.54 & -0.136 \\
7 & 11487828 & $6.82\cdot 10^{9}$ & 7.20 & -0.014 \\
8 & 6031050 & $3.68\cdot 10^{9}$ & 8.88 & -0.096 \\
9 & 5443067 & $3.29\cdot 10^{9}$ & 8.62 & -0.187 \\
\hline
\end{tabular}
\end{table}

\bsp	
\label{lastpage}
\end{document}